\documentclass[mathleft]{an}

\makeatletter
\let\NAT@parse\undefined
\makeatother

\usepackage{times}
\usepackage[]{natbib}
\usepackage[utf8]{inputenc}
\usepackage{graphicx}
\usepackage{times}
\newcommand{\rj}{R$_j$}
\begin{document}

\Pagespan{1}{1}
\Yearpublication{}%
\Yearsubmission{}%
\Month{}%
\Volume{}%
\Issue{}%
\title{
Interaction of radio jets with clouds in the ambient medium: Numerical simulations
}

\author{Solai Jeyakumar\fnmsep\thanks{Corresponding author:
  \email{sjk@astro.ugto.mx}\newline}
}

\titlerunning{Radio jets: Numerical simulations}
\authorrunning{S. Jeyakumar}
\institute{
Departamento de Astronomía, Universidad de Guanajuato, AP 144, Guanajuato CP 36000, Mexico. 
}

\received{}
\accepted{}
\publonline{later}


\abstract {%
Hydrodynamical simulations of jets interacting with clouds moving in the ambient
medium of the host galaxy are presented. Clouds with sizes of the order of the 
jet diameter and smaller, crossing the path of the jet with different speeds are considered. 
In the case of slow moving clouds the jet is  stopped over the brief period of time 
taken by the cloud to cross the jet. The jet maintains its general morphology 
in the case of fast moving clouds. Erosion of the clouds leads to redistribution
of cloud material to large distances. Such interaction may explain the 
large outflow velocities observed from pc to kpc scales in the compact radio sources.
}

\maketitle
\section{Introduction}
The Gigahertz Peaked Spectrum (GPS) and the Compact
Steep Spectrum (CSS) sources \citep[cf.][for a review]{odea98}
are young radio sources evolving in the Inter Stellar Medium (ISM) 
of the galaxies hosting them.  The radio morphology of these sources show 
evidence of interaction of the radio jets with  inhomogeneties in the ISM.
\citep{saikia.etal95, sjk.etal05}. 
Several observations of the line emitting gas in the optical and radio wavelengths, 
show further evidence for such interactions. Observations of the gas at 
large scales reveal that the line emission from the 
narrow line regions (NLRs) and the cold neutral HI clouds show 
complex velocity profiles and outflows 
\citep{devries.etal99, odea.etal03, labiano08,labiano.etal05, emonts.etal05, morganti.etal05}.
In addition, outflows have been observed in the broad components. 
These observations reveal outflows of line emitting gas
with velocities ranging from a few 100 km/s to more than 1000 km/s, from pc to kpc scales, 
both in the ionised and the neutral components 
(cf. Holt, Tadhunter \& Morganti 2003, 2008; Morganti et al. 2005).
Modelling of lines of high ionisation state (Gupta, Srianand \& Saikia 2005),
as well as several narrow lines \citep{labiano.etal05}, suggest 
that there is a contribution by shocks to the ionisation of the gas. 
All these observations suggest that interaction of jets with clouds
in the ISM could play a role in the ionisation and outflow of the gas.

Numerical simulations have been successfully used to study 
the effects of propagation of radio jets through a variety of
ambient atmospheres \citep[cf.][]{wiita&norman92, hooda&wiita96,sjk.etal05,carvalho&odea02b, krause05}. 
Such studies can be used to understand the relation between
the outflow and the ionisation of the gas. In fact, several numerical studies 
have been attempted to study the interaction of the jets with clouds in 
the ambient medium \citep{higgins.etal99, wang.etal00,xu&stone95,fragile.etal04,
choi.etal07}.  All these studies focused on the evolution of the individual clouds and
the effect of the clouds on the dynamics and morphology of the radio source.
The clouds in these studies are large and massive resembling diffuse neutral 
clouds in the ISM of the Galaxy. A few simulations of jet propagation in a clumpy medium
have  been carried out where a few of them also considered parameters similar 
to that of the CSS/GPS sources, but with a lower jet kinetic power
\citep{saxton.etal05, steffen.etal97, sutherland&bicknell07}. 

In all of these simulations, the jet collides with a static cloud or
a dense clumpy structure in the ambient medium.  However, a more realistic scenario 
is dense small  clouds such as those expected in the NLR and BLRs that move into 
to the path of the jet/lobe. Such a scenario is more probable for the young
radio sources where the radio jet has not cleared the natal cocoon of material 
of the ISM yet. So, here simulations of clouds moving into the jet are attempted.

\begin{figure*}[t]
\vbox{

\hbox{
\includegraphics[width=5.5cm,clip=]{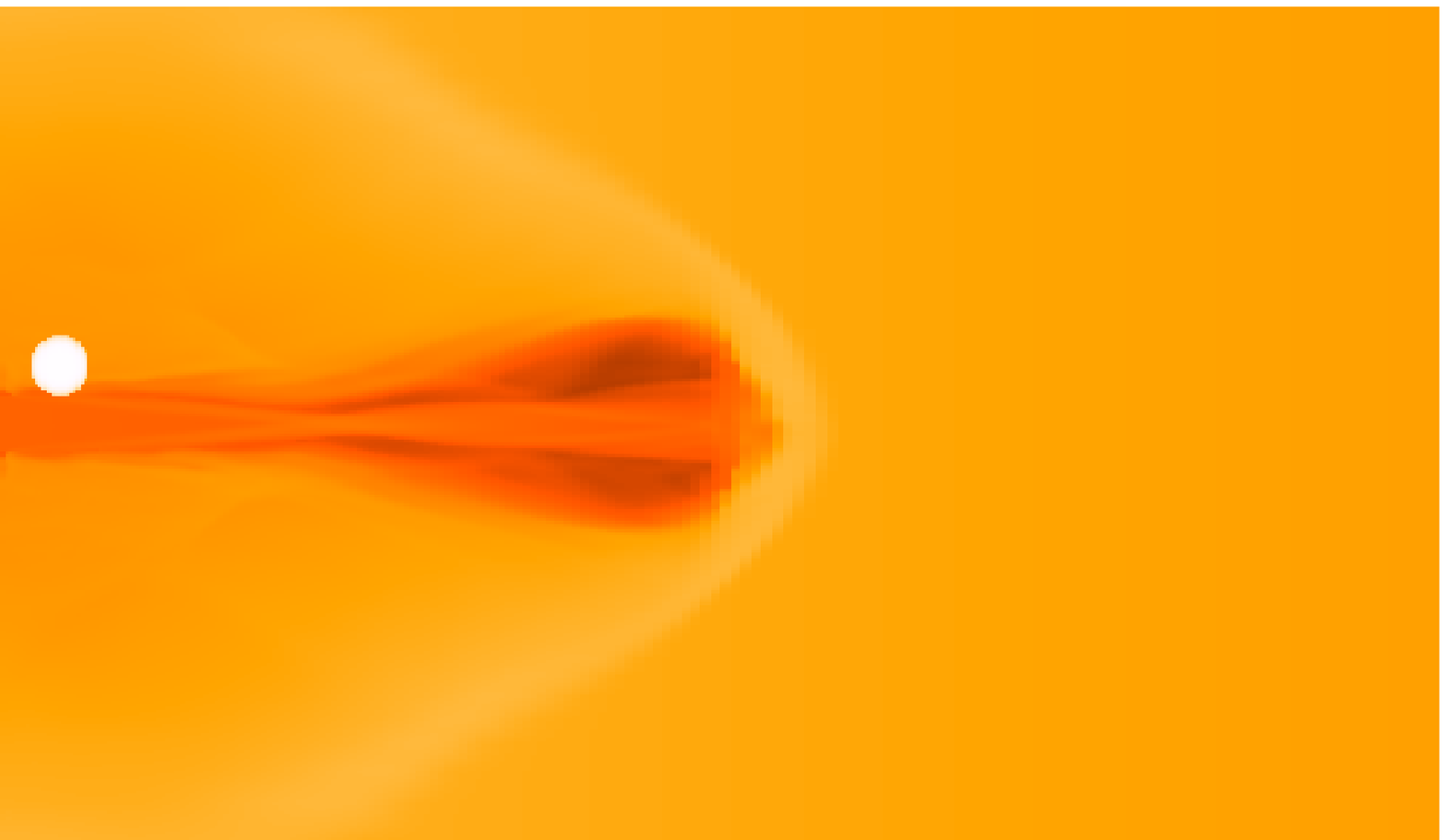}
\includegraphics[width=5.5cm,clip=]{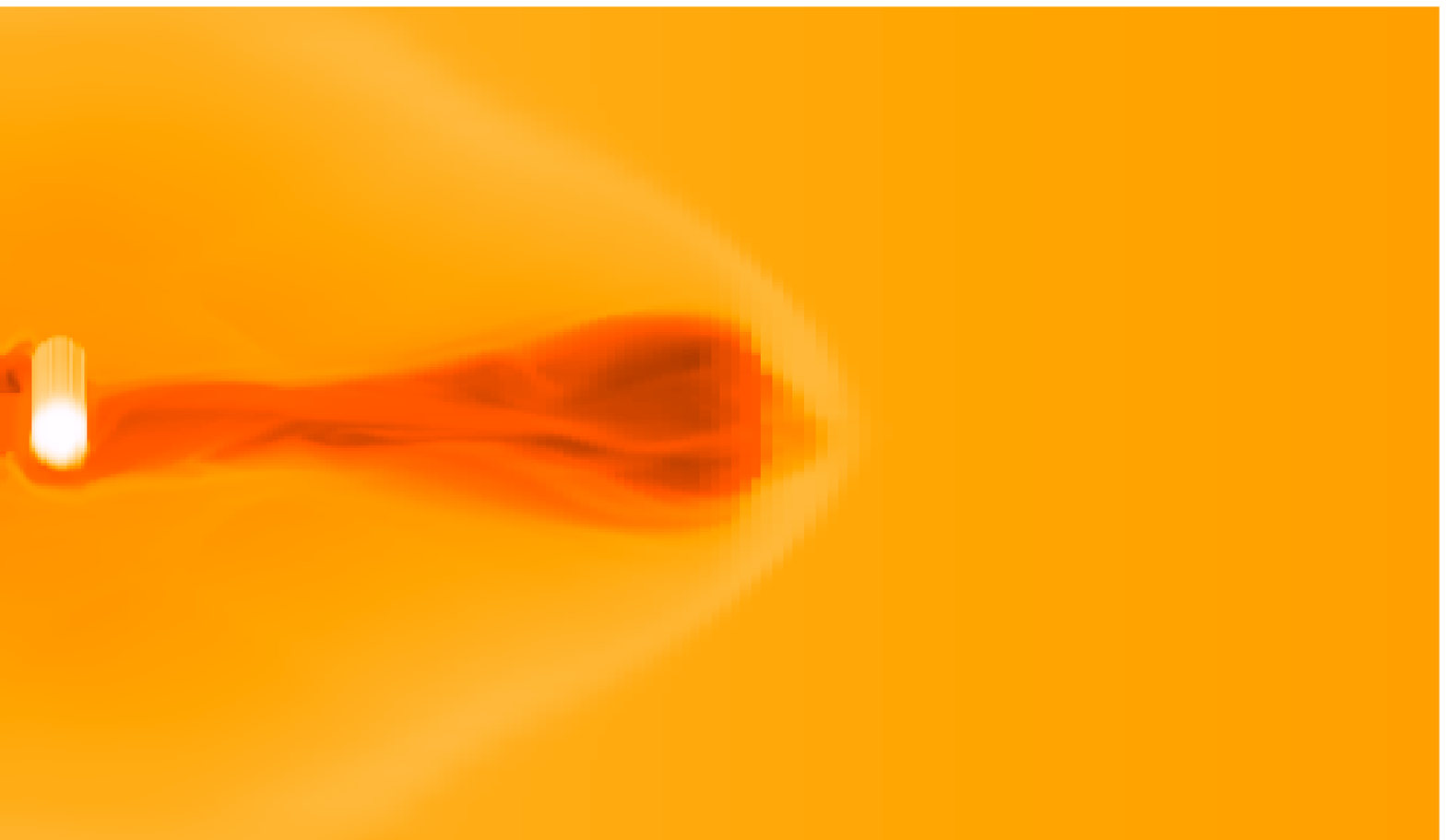}
\includegraphics[width=5.5cm,clip=]{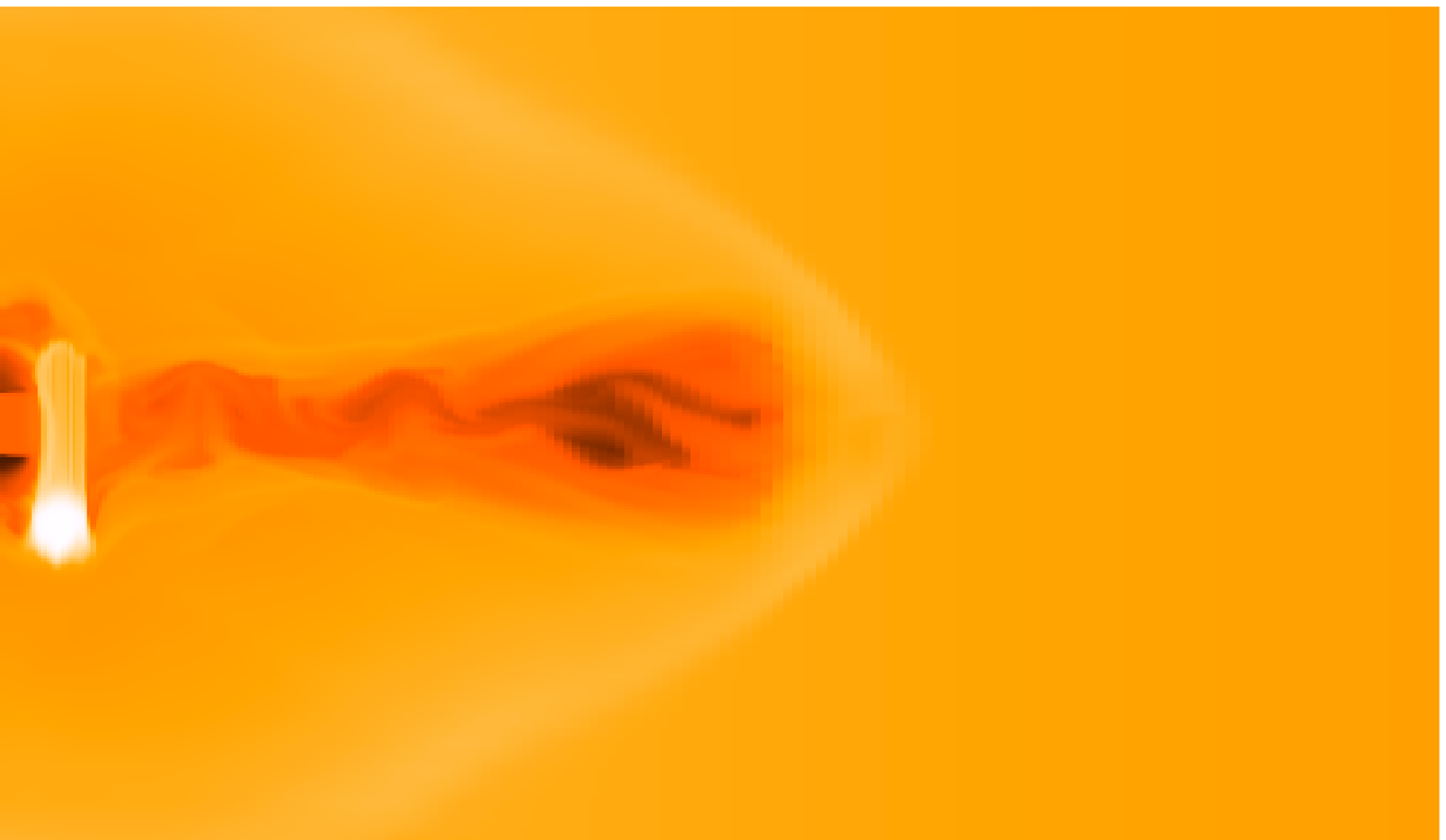}
}

\hbox{
\framebox[5.5cm]{
\hspace*{.1mm}
\includegraphics[width=5.5cm,clip=]{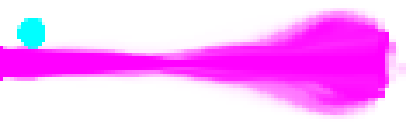}
}
\framebox[5.5cm]{
\hspace*{.1mm}
\includegraphics[width=5.5cm,clip=]{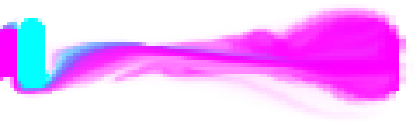}
}
\framebox[5.5cm]{
\hspace*{.1mm}
\includegraphics[width=5.5cm,clip=]{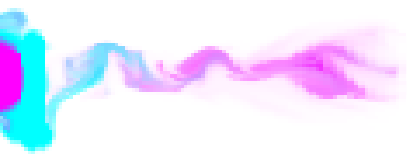}
}
}

}
\caption{Snapshots of the logarithm of density (top) and the Lagrangian tracers 
of the cloud and jet material (bottom) of the run F1 are shown. 
The images from left to right correspond to 1.0, 1.04 and 1.12 problem time units.
Cyan color (light gray) represents the cloud material and magenta (dark gray) 
represents the jet material. The radius of the jet and the size of the cloud are scaled to 100~pc.
In this unit the total extent of the visible jet is about 3~kpc (left panel). 
\label{fig:sim1}
}
\end{figure*}

\begin{figure*}
\vbox{

\hbox{
\includegraphics[width=5.5cm,clip=]{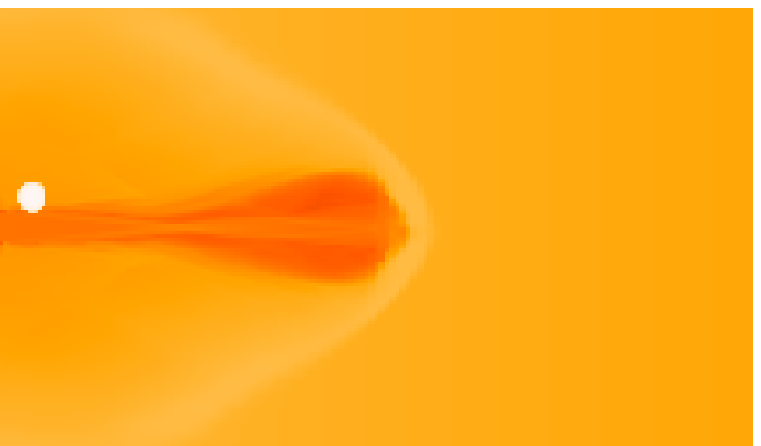}
\includegraphics[width=5.5cm,clip=]{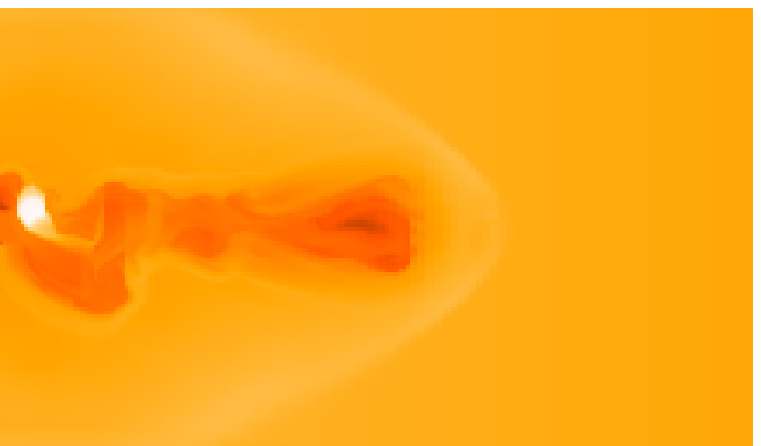}
\includegraphics[width=5.5cm,clip=]{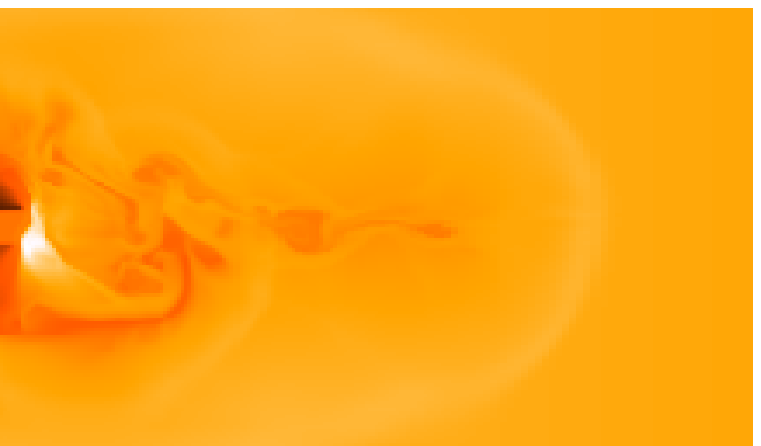}
}

\hbox{
\framebox[5.5cm]{
\hspace*{.1mm}
\includegraphics[width=5.5cm,clip=]{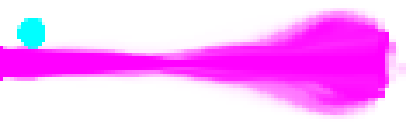}
}
\framebox[5.5cm]{
\hspace*{.1mm}
\includegraphics[width=5.5cm,clip=]{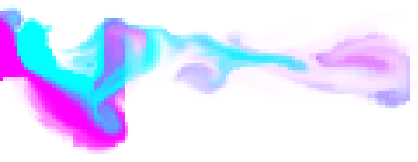}
}
\framebox[5.5cm]{
\hspace*{.1mm}
\includegraphics[width=5.5cm,clip=]{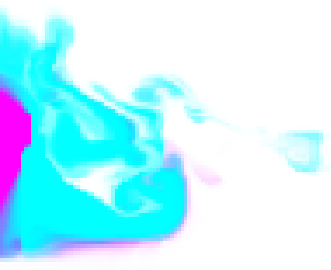}
}
}

}
\caption{Same as Fig~\ref{fig:sim1}, but for the run S1 (slow cloud). 
The snapshot times are 1.0, 1.18 and 1.65 problem time units from left to right.
The radius of the jet is scaled to 100~pc and the size of the cloud is 100~pc.
In this unit the total extent of the visible jet is about 3~kpc (left panel). 
\label{fig:sim2}
}
\end{figure*}

\begin{figure*}
\vbox{

\hbox{
\includegraphics[width=5.5cm,clip=]{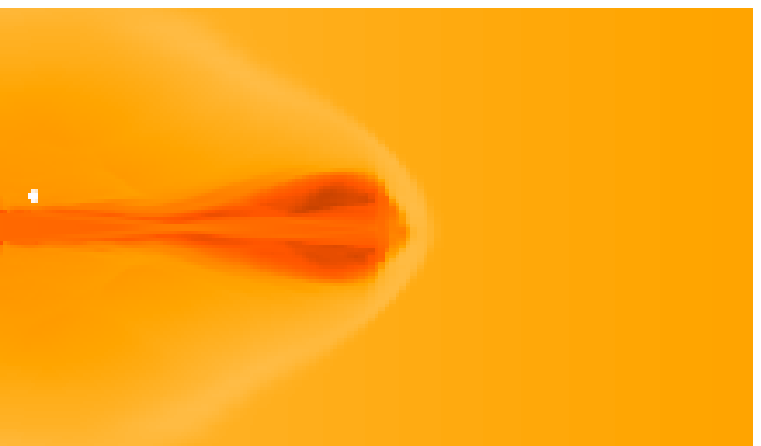}
\includegraphics[width=5.5cm,clip=]{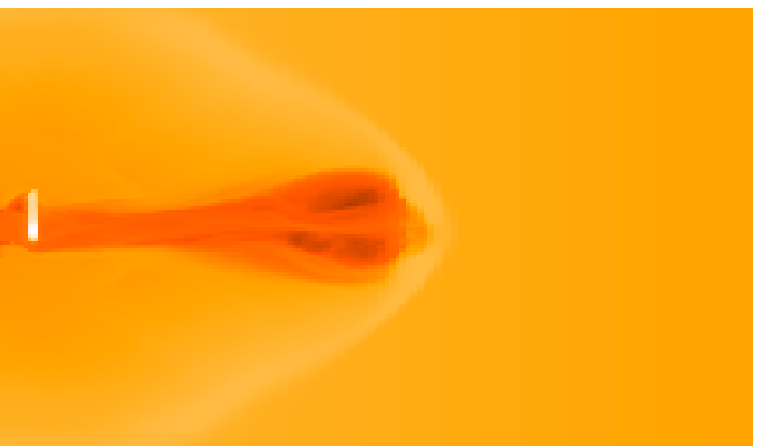}
\includegraphics[width=5.5cm,clip=]{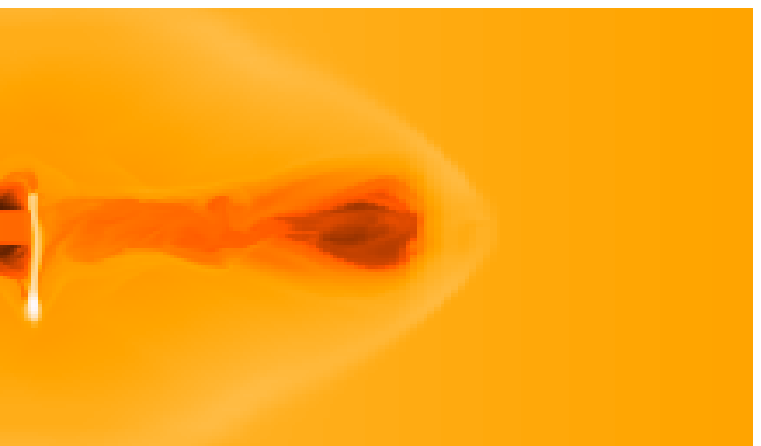}
}

\hbox{
\framebox[5.5cm]{
\hspace*{.1mm}
\includegraphics[width=5.5cm,clip=]{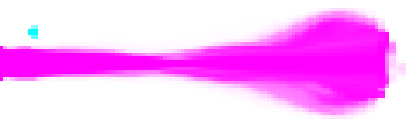}
}
\framebox[5.5cm]{
\hspace*{.1mm}
\includegraphics[width=5.5cm,clip=]{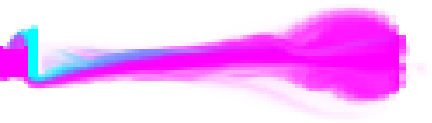}
}
\framebox[5.5cm]{
\hspace*{.1mm}
\includegraphics[width=5.5cm,clip=]{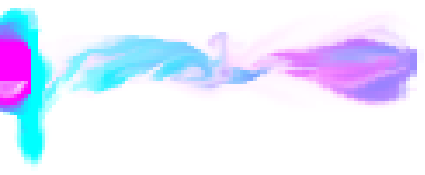}
}
}

}
\caption{Same as Fig~\ref{fig:sim1}, but for the run F2.
The radius of the jet is scaled to 100~pc and the size of the cloud is 50~pc.
\label{fig:sim3}
}
\end{figure*}

\begin{figure*}
\vbox{

\hbox{
\includegraphics[width=5.5cm,clip=]{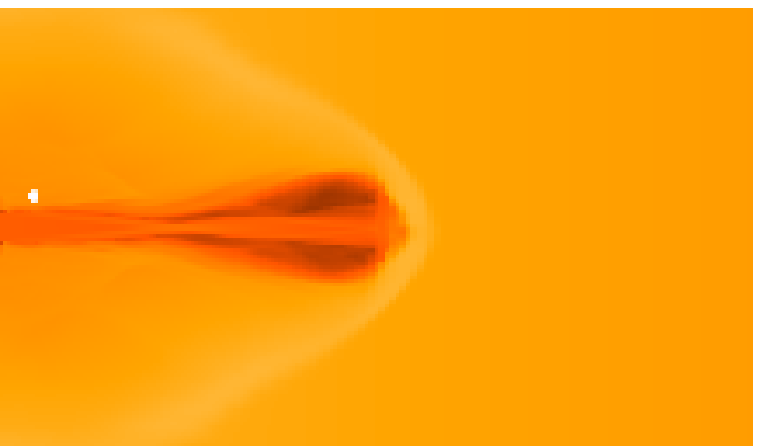}
\includegraphics[width=5.5cm,clip=]{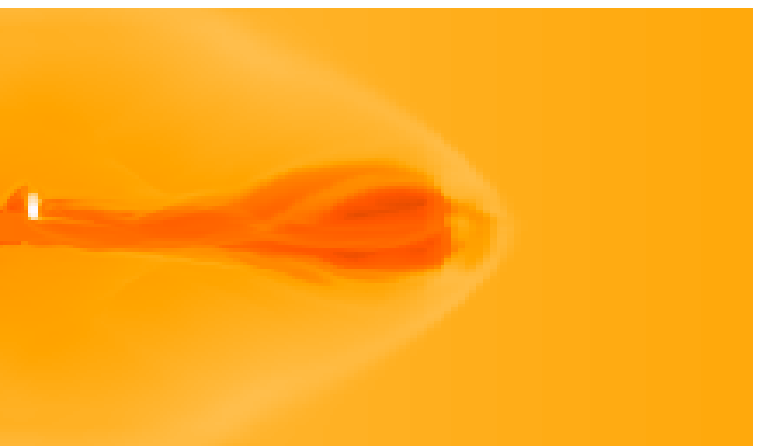}
\includegraphics[width=5.5cm,clip=]{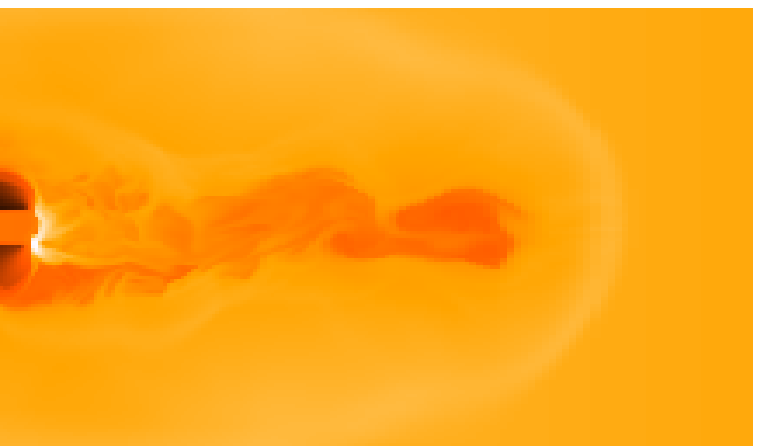}
}

\hbox{
\framebox[5.5cm]{
\hspace*{.1mm}
\includegraphics[width=5.5cm,clip=]{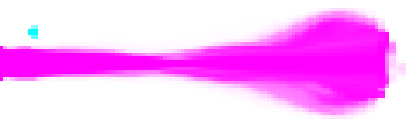}
}
\framebox[5.5cm]{
\hspace*{.1mm}
\includegraphics[width=5.5cm,clip=]{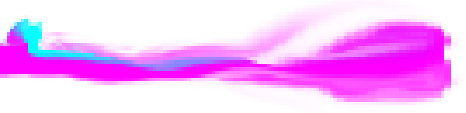}
}
\framebox[5.5cm]{
\hspace*{.1mm}
\includegraphics[width=5.5cm,clip=]{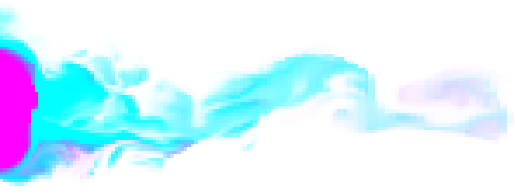}
}
}

}
\caption{Same as Fig~\ref{fig:sim2}, but for the run S2.
The radius of the jet is scaled to 100~pc and the size of the cloud is 50~pc.
\label{fig:sim4}
}
\end{figure*}

\section{Jet cloud interaction}
The line emitting regions of the ISM are expected to be clumpy and turbulent. 
It can be approximated as a hot tenuous inter-clump
medium with spherical clumps embedded in it. There are two ranges
of densities and scale sizes of the clumps in the ISM that show outflows: 
(a) the NLRs and dense large neutral clouds 
($n~\sim~10^{2-5}$~cm$^{-3}$) at kpc scales and, (b) BLRs with very 
high densities ($n~\sim~10^{11}$cm$^{-3}$) but at sub-pc scales 
\citep[cf.][]{ferland.etal92, xu.etal07, sulentic.etal00, laor.etal97}. 
To simulate a cloud with a typical density of BLR, a grid resolution 
of milli parsec is required. For the purpose of studying the 
overall dynamics involved in the interaction of the jet with a moving cloud,
small dense clouds at two different velocities are considered.
The density of the clouds considered here is comparable to that of the NLRs. 
The sizes of the clouds are chosen such that at least 10 zones span the cloud diameter. 
Because of this constraint the clouds are very massive. A smaller cloud 
spanning only 5 zones are also considered to reduce the mass of the cloud.

\subsection{Numerical setup}
Hydro-dynamical simulations in 3D were carried out using the ZEUSMP-V2.0  code,
making use of the  parallel feature that has become available with this code
\citep{hayes.etal06,stone&norman92a,stone&norman92b, clarke96a}.
The jet problem was initialised based 
on the previous jet launching code \citep[cf.][]{sjk.etal05}. 
For the simulations presented here a cluster of  20 processing cores were used. 
These simulations were run on a Cartesian grid with $180\times126\times22$ active zones 
spanning 50~\rj\ in the jet propagation direction, $x$,
and $\pm$15~\rj\ in the cloud propagation direction, $y$, and $\pm1.1$~\rj\  in the $z$ direction, where \rj\ is the radius of the jet.
This choice allows us to consider a better resolution in two directions at the cost
of a smaller physical extension in the third direction, although the resolution is the same. 
In the $y$ direction, 80 uniform zones were used to span the inner 8~\rj\ region, 
while the rest of the zones were logarithmically scaled. In the $z$ direction all
zones were uniformly spaced. In the $x$-direction logarithmic scaling was used.
This enables us to track the propagation of the cloud with a reasonable resolution.
Outflow boundary conditions were used everywhere except at the jet inlet, where 
inflow boundary condition was used.

The unit of length in this problem is the radius of the jet, \rj. The 
unit of time is given by the sound speed of the ambient medium, which was set to 1.0.
The important parameters of the jet are
the intrinsic Mach number, $M_j$, and the density, $\eta$, with respect
to the ambient medium density. 
The initially pressure matched jet is assumed to be conical with a half opening angle of $\Omega$.
The ambient density is given as, $n_a(r) = n_0/ ( 1 + (r/a)^2)^{\delta}$,  where $a$ is the 
core radius, $\delta$ is the power law index of the density profile and $r$ is the distance from the nucleus of
the source. 

The cloud is characterised by the central density, $n_c$,  size, $R_c$, and a velocity
with respect to the sound speed of the ambient medium, $V_c$. The density of the 
cloud is tapered as $n_a(r) +  (n_c - n_a(r)) \tanh( (R_c - R)/(0.3~R_c) )$, to avoid numerical artifacts 
(cf. Choi et al. 2007), where $R$ is the distance from the centre of the cloud. 
The temperature of the cloud is varied in such a way that the pressure inside follows
the pressure of the undisturbed ambient medium, although the cloud is introduced 
inside the region affected by the bow shock of the jet, leading to a small but negligible
pressure difference between the cloud and the disturbed ambient medium.
In addition, Lagrangian tracer variables are used to track the jet and 
the cloud materials using the multi-species advection scheme available 
with ZEUSMP V2.0 (see Hayes et al. (2006), for more details). These variables are set
as 1.0 at the corresponding grid locations at the initial times.

The numerical parameters used in these simulations are,
$M_j=26$, $\eta=0.001$, $a=20$~\rj, $n_0=1.0$,  $\delta=0.75$ and $\Omega=0.02$.
After the jet has propagated to 
about 30~\rj, the cloud is introduced at $x$=2~\rj, $y$=2~\rj\  and $z$=0. 
The numerical runs  F1 and F2 (called fast cloud) with an $R_c$ of 1.0 and 0.5~\rj\  
respectively are calculated with a velocity $V_c$ of 50. The numerical runs S1 and S2 
(called slow cloud) are calculated  with the same parameters but with a  $V_c$ of 5. 

These simulations are scaled as follows: for a jet radius of 100~pc,
and  temperature of the ambient medium of $2\times10^6$ K, the time unit 
is 0.43 Myr and the jet power is $2\times10^{45}$~erg/s.
For these parameters the velocities of the fast and slow clouds 
are 10$^4$ and 10$^3$ km/s and the core radius of the 
density profile of the ambient medium is 2~kpc. The central density of the cloud is
10$^3$~cm$^{-3}$.

\section{Results and Discussion}
Color images of the logarithm of the density as well as the Lagrangian tracers of the
jet and the cloud are shown in Fig~\ref{fig:sim1} to Fig~\ref{fig:sim4}.
These figures show the morphology of the jet and the cloud at different times: 
(a) when the cloud is launched,  (b) when the cloud is half-way through
the jet and (c) when the cloud has just passed through the jet. During the initial
interaction of the cloud with the jet, erosion of the cloud material is seen
(see the centre panels of the figures). The eroded cloud material is
carried to large distances. Apart from this, the cloud remains as a clump
blocking the path of the jet. At later stages, distortion of the cloud can be seen
and the cloud material is distributed to a distance of about half
the length of the lobe.  At later stages when most part of the cloud has passed
through the jet, a clump of material can be identified as a cloud, but compressed
in the direction of propagation of the jet. 

The time taken by the slow, small cloud to travel a distance of a jet diameter is 
about $3\times10^5$ yr, while the larger cloud takes somewhat longer time 
to cross through the jet.  The jet is stopped completely in the slow cloud runs
and the morphology is similar to that seen in other simulations of static clouds.
In the fast cloud simulations, the cloud velocity is 10 times faster, 
thus the interaction times are smaller. These simulations also show the general morphology 
seen in the slow cloud runs. However, the jet is not completely stopped in these runs
and the radio source maintains its general morphology. Although 
the parameters of the jet and those of the clouds used here are 
somewhat different between the simulations of interaction of the jets with 
static clouds (e.g. Higgins et al. 1999; Wang et al. 2000), these results 
can be compared with that of the static cloud simulations. The main difference
is the erosion of the cloud material at the initial stages. 
Since the clouds are very massive, the clouds do not gain any velocity 
in the $x$ direction. However, the eroded material seen at large distances 
during the initial encounter accelerates to large distances and can be 
compared with the outflow. At later times (the right panels of the figures), 
back flowing cocoon also carries the material backwards. In such a scenario 
the emission lines can be very complex. 

In young radio sources where the radio jets are still expanding within the 
natal cocoon of a clumpy ISM, such interactions are likely to occur. 
Once the jet has emerged out of the ISM clearing the cocoon of material on its way,
such incidences are less likely to occur. Observations also indicate
such a scenario \citep{inskip.etal02}. These simulations also suggest
that the distorted radio morphology and  outflow can be correlated. 


\acknowledgements
The work presented here was supported by PROMEP/103-5/07/2462 grant.
I thank the referee Chris O'Dea for very useful comments.

\bibliographystyle{aa}
\bibliography{radio}



\end{document}